\documentclass[%
 reprint,
 amsmath,amssymb,
 aps,
]{revtex4-2}

\usepackage{graphicx}
\usepackage{dcolumn}
\usepackage{bm}

\begin{document}

\preprint{APS/123-QED}

\title{Quantifying Emergent Behaviors in Agent-Based Models \\using Mean Information Gain}

\author{Sebastián Rodríguez-Falcón}
\affiliation{%
Pontificia Universidad Católica del Perú, Lima, Perú
}%

\author{Luciano Stucchi}
\affiliation{
 Universidad del Pacífico, Lima, Perú
}%

\date{\today}

\begin{abstract}
Emergent behaviors are a defining feature of complex systems, yet their quantitative characterization remains an open challenge, as traditional classifications rely mainly on visual inspection of spatio-temporal patterns. In this Letter, we propose using the Mean Information Gain (MIG) as a metric to quantify emergence in Agent-Based Models. The MIG is a conditional entropy–based metric that quantifies the lack of information about other elements in a structure given certain known properties. We apply it to a multi-agent biased random walk that reproduces Wolfram’s four behavioral classes and show that MIG differentiates these behaviors. This metric reconnects the analysis of emergent behaviors with the classical notions of order, disorder, and entropy, thereby enabling the quantitative classification of regimes as convergent, periodic, complex, and chaotic. This approach overcomes the ambiguity of qualitative inspection near regime boundaries, particularly in large systems, and provides a compact, extensible framework for identifying and comparing emergent behaviors in complex systems.


\end{abstract}

\maketitle

Complex systems are composed of numerous interconnected and interdependent components \cite{fromm2004emergence}, and while their components may be similar, behaviors emerge from interactions between them rather than individual parts themselves. Unlike reductionist scientific approaches that study components in isolation \cite{siegenfeld2019introduction}, complex system science examines how elements within a system influence each other. Hence, complex systems exhibit self-organization, adaptation, and emergence \cite{Vicsek2012}, which means that they develop structured patterns without central control. These emerging behaviors that arise from the interactions of individual components \cite{Goldenfeld1999} can be observed in physical, biological, or social systems \cite{siegenfeld2019introduction}. For example, water molecules remain the same across different phases, yet their properties vary \cite{Anderson1972} as a result of differences in molecular interactions. Similarly, large-scale behaviors such as flocking birds \cite{Couzin2005}, decentralized economies \cite{Tesfatsion2002}, and evolving ecosystems emerge through self-organization \cite{Levin1998} and cannot be understood by examining individual components alone. These kinds of phenomena are considered emergent, and emergence is recognized as a fundamental characteristic of complex adaptive systems \cite{artime2022emergent} and a key principle in disciplines such as biology (nervous and immune systems), neuroscience (neural networks), social sciences (social networks, cultures and languages), systems ecology, and economics \cite{siegenfeld2019introduction}. Although human influence shapes technological and cultural systems, many natural systems organize themselves through internal interactions between their components \cite{fromm2004emergence}.

A common approach to studying complex systems is through computer simulations \cite{fromm2004emergence}, in which simple local rules are repeatedly applied to interacting agents, revealing the intricate dynamics of the system as a whole. Among the available modeling approaches, Agent-Based Modeling (ABM) and Cellular Automata (CA) have proven to be particularly suitable \cite{Bonabeau2002} for studying emergent phenomena. CA are discrete, spatio-temporal dynamic systems governed by local rules. They consist of four key elements \cite{wolfram1983statistical}: a grid of cells with finite states, a neighborhood defining interaction, initial conditions assigning states to each cell, and rules that update cell states based on neighboring cells. The model evolves by iteratively applying these rules and updating the grid at each step. ABM is a computational approach used to simulate the behaviors and interactions of autonomous agents within a system \cite{wilensky2015abm}, such as individuals, organizations or entities. Each agent operates independently, attempting to achieve specific goals, and follows simple rules that determine its interactions with others. Unlike CA, which primarily examines large-scale patterns resulting from local interactions, ABM focuses on understanding how variations in agent characteristics, decision-making processes, or rule changes affect system dynamics \cite{Grimm2005}. Consequently, while both CA and ABM can be used as simple models for complex behaviors, ABM further enables a qualitative description of the underlying dynamics of the complex phenomenon.\\

A central challenge in studying complex systems is characterizing and quantifying their emergent behaviors \cite{Mitchell2009}. These behaviors often manifest themselves as distinct visual patterns \cite{Grimm2005} in one-, two-, or three-dimensional spaces. Efforts to classify these behaviors date back decades \cite{Goldstein01031999}, and it is customary to use a four-class scheme, distinguishing between the convergent, periodic, chaotic, and complex regimes proposed by Stephen Wolfram \cite{wolfram2020computation}. However, much of the field has historically relied on system-specific metrics \cite{rosas2020reconciling}. Although previous work has examined both the quantitative \cite{langton1986} and qualitative \cite{wolfram1983statistical, wolfram1984universality, Wolfram2002, halley2008classification} emergent behavior on cellular automata, it has mainly aimed to classify their rule spaces and computational characteristics with limited progress toward generalizable quantitative measures that can capture the essence of emergence across different models. This is where the challenge lies in studying emergent phenomena: we can see the patterns emerging from the interactions within the system, but we still lack a formal, comprehensive approach to fully understand and mathematically describe how these patterns emerge, leaving a critical problem in the study of complex systems unresolved.\\

In this Letter, we propose a method to quantify differences in system behaviors using a modified conditional entropy metric, the Mean Information Gain (MIG) derived from the Information Gain defined by \cite{andrienko2000complexity,orderchaosandcomplexity,WACKERBAUER1994133}. Motivated by Parrott’s use of MIG to quantify ecological complexity \cite{PARROTT20101069}, we show how this metric connects the emergence to classical notions of order and disorder in physics. This is illustrated through the application of the MIG to a multi-agent biased random walk in a two-dimensional discrete space, a toy-model capable of reproducing all four behavioral regimes described by Wolfram by just varying the values of its parameters.\\

\begin{figure}
\includegraphics[scale=0.25]{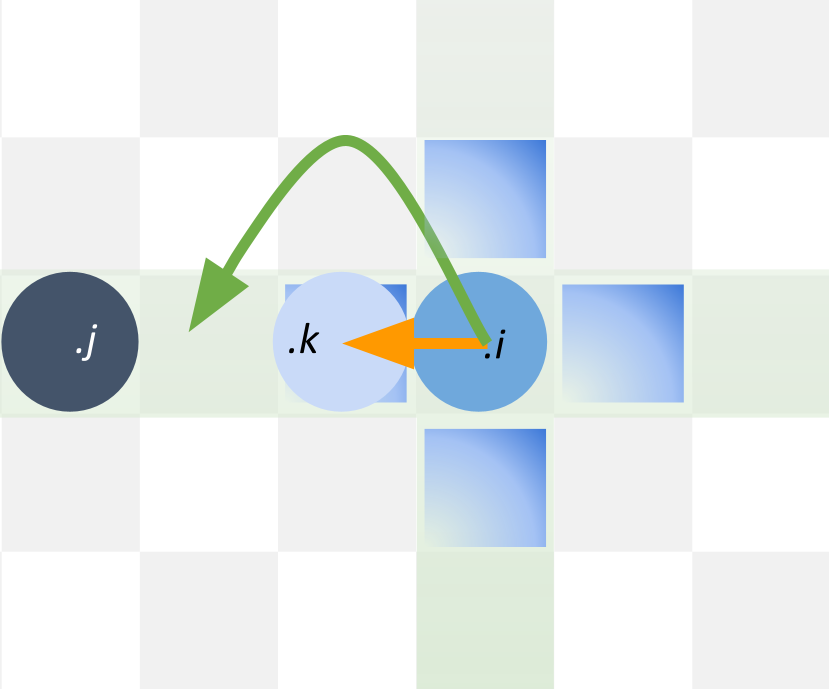}
\caption{\label{wide:vision}
Illustration of agent perception and movement based on the \textit{vision} and \textit{superposition} parameters. The green cells define the \textit{orthogonal} vision range, while the blue cells correspond to the \textit{Von Neumann} neighborhood. In this example, agent \textit{i} has orthogonal vision and selects agent \textit{j} as its target. If \textit{superposition} is enabled, agent \textit{i} takes one step in \textit{j} direction and moves directly onto agent \textit{k}'s cell, as indicated by the orange arrow. If \textit{superposition} is disabled, agent \textit{i} moves to the nearest available cell in the direction of agent \textit{j}, as indicated by the green arrow.}
\end{figure}

\textit{Model.--} We implemented a multi-agent biased random walk in a two-dimensional discrete space. In this rule-based computational model, agents are randomly distributed within a room and the biased movement is implemented as a simple two-step rule: first, randomly select another agent within their field of view, and then take a single step towards them. If no such agents are nearby, the next step is taken randomly. The parameters that shape the dynamics of agents with their environment are the \textit{vision} and \textit{superposition}. The agent's \textit{vision} values are the \textit{Von Neumann} vicinity or the \textit{orthogonal} vicinity. The \textit{superposition} property determines whether the agents are able to share the same cell or not. The movement dynamics and considerations are explained in Fig. \ref{wide:vision}.  \\

\begin{figure}
\includegraphics[scale=0.27]{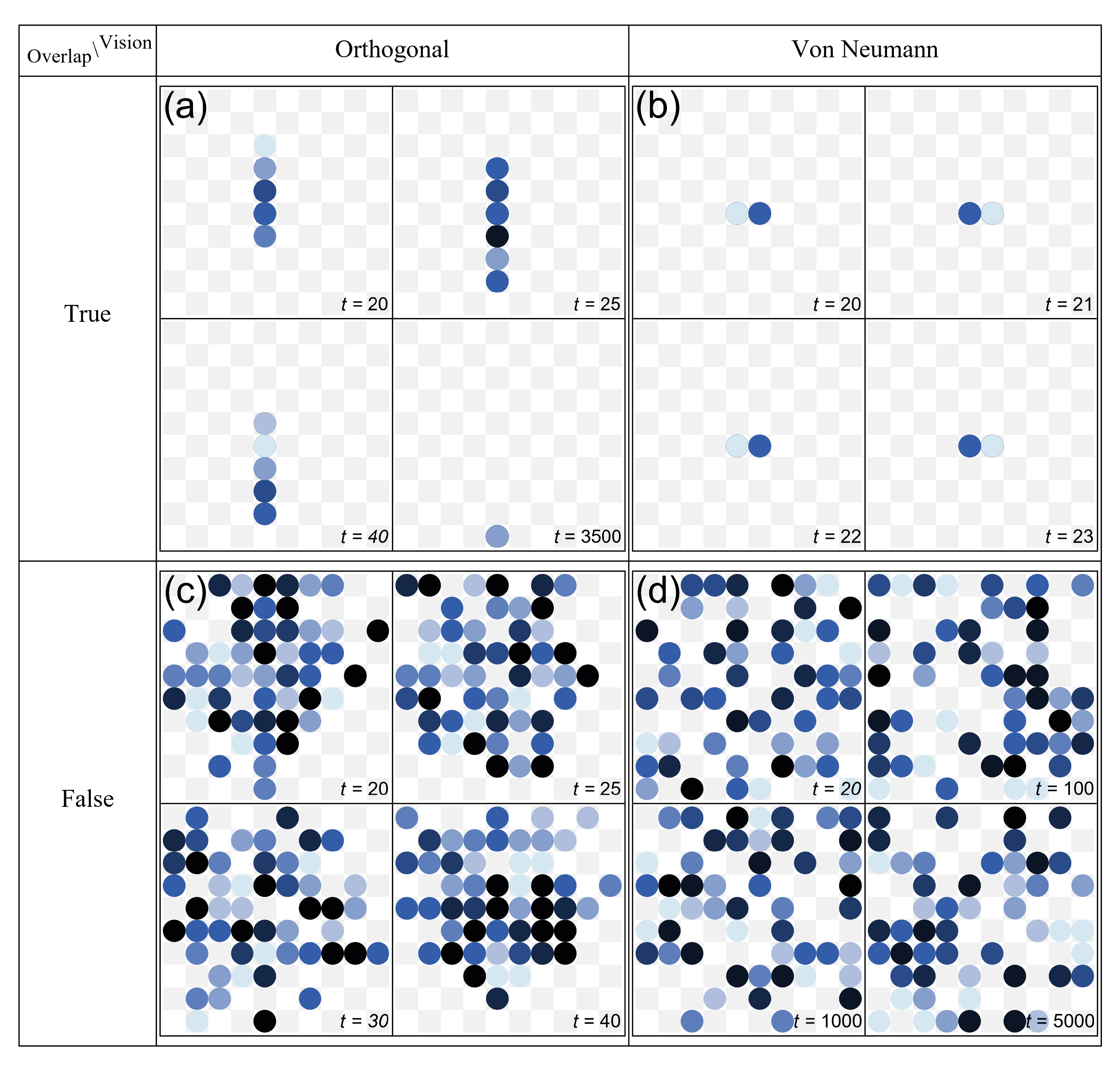}
\caption{\label{wide:emergent} Emergent phenomena for each system configuration. (a) Convergent regime: Initially oscillates in one direction until it converges in the border. (b) Periodic regime: Quickly converges into oscillating clusters, in some cases they all converge to only one. (c) Complex regime: Present a close-to-chaotic behavior but agents tend to move together. (d) Chaotic regime: Does not present any recognizable pattern.}
\end{figure}

\textit{Emergent Phenomena.--} Systematically exploring all possible combinations of the parameters of our model, we obtain four qualitatively distinct emergent behaviors, which correspond to the complete set of regime types described in Wolfram’s classification, as illustrated in Fig. \ref{wide:emergent}. However, the classification based solely on visual inspection remains subjective and qualitative. To objectively characterize the collective behavior of the system and its evolution, a quantitative measure is required.\\

\textit{Quantifying Emergence.--} As a complexity measure, we introduce $\bar{G}$ as the MIG. According to \cite{andrienko2000complexity}, it quantifies the lack of information on other elements of a structure, given that certain properties of the structure are known. It is expressed as
\begin{equation}
\bar{G}_{X,Y}=-\sum_{x,y}P(x,y)\log_{2}P(x|y)
\end{equation}

where X and Y are discrete random variables with \( x \in X \) and \( y \in Y \). $P(x|y)$ is the conditional probability of a state \textit{x} conditioned on the state \textit{y} and $P(x, y)$ is the joint probability. $\bar{G}$ quantifies the average information gain (measured in bits) on the value $x$ given the knowledge of the value $y$ \cite{javaheri2016information}. \\ 

In the context of our model, the structural complexity $G$ of a given configuration is the total MIG of cells having a homogeneous and heterogeneous neighborhood \cite{javaheri2016information}. To apply this metric to a multi-agent biased random walk in a two-dimensional discrete space, we assign each cell a binary state: 0 if unoccupied and 1 if occupied by at least one agent. By registering each cell state per unit time, the variables can be redefined as follows:

\begin{equation}
\overline{G}_{s_r, s_\Delta{r}}=-\sum_{s_{r},s_{\Delta{r}}}P(s_r,s_{\Delta{r}})\log_{2}P(s_r|s_{\Delta{r}})
\end{equation}

where $s_r$ will be the state (0 or 1) of the reference agent and $s_{\Delta{r}}$ indicates the state of the agent with position $\Delta{r}$ relative to the reference agent. In this case, we consider the direction $\Delta{r}$ to be up, down, left, and right. \\

\textit{Experiment.--} The agent-based model was implemented in NetLogo \cite{Wilensky:1999}. For each combination of the parameters (\textit{vision} and \textit{superposition}), we ran experiments under the following conditions for each regime: 

\begin{itemize}
    \item Convergent: 100 repetitions, 20,000 time steps.
    \item Periodic: 1000 repetitions, 5000 time steps.
    \item Complex and Chaotic: 100 repetitions, 1000 time steps.
\end{itemize}

During each simulation, the positions of all agents were recorded at each time step to analyze the spatial evolution of the system. Different time-step horizons were used for each regime to account for their distinct dynamical timescales. The convergent regime required a longer simulation to allow all agents to fully collapse into a single point, while the periodic regime needed time for agents to gather and establish sustained oscillations. In contrast, complex and chaotic regimes reach their characteristic behaviors rapidly; therefore, we evaluated them only over 1000 time steps to analyze their evolution. To compute the MIG, we used positional data to determine whether each cell was occupied or empty at each time step. For every type of regime, we calculated the MIG in each of the four directions (up, down, left, right), then averaged the results over time and across all repetitions. \\

\begin{figure}
\includegraphics[scale=0.28]{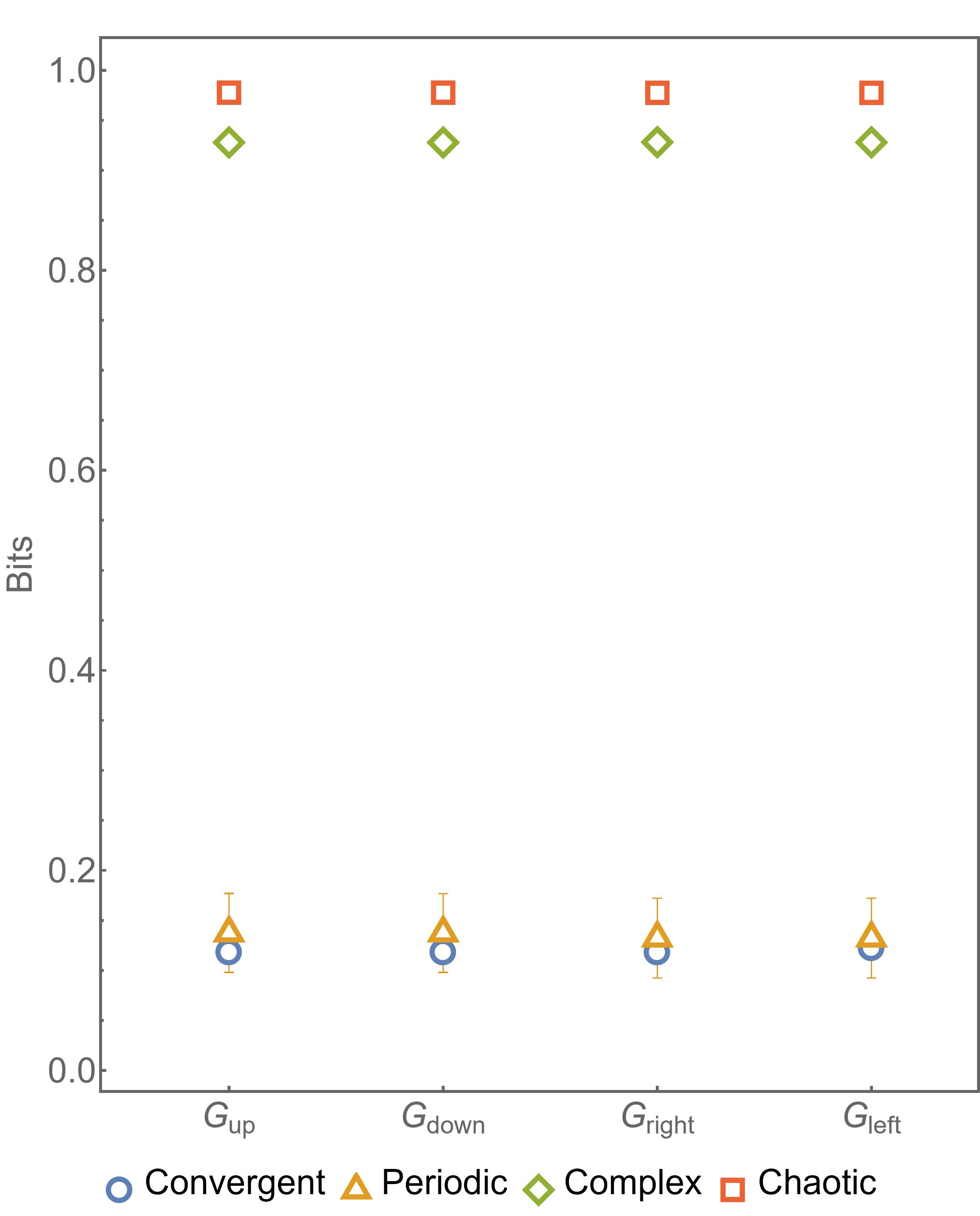}
\caption{\label{wide:gconv} 
MIG per direction, averaged over time and across all repetitions, for each behavioral regime. 
Convergent: $G_{up}=0.1181 \pm 0.0024$, $G_{down}=0.1182 \pm 0.0025$, $G_{right}=0.1180 \pm 0.0040$, $G_{left}=0.1220 \pm 0.0040$; 
Periodic: $G_{up}=0.14 \pm 0.04$, $G_{down}=0.14 \pm 0.04$, $G_{right}=0.13 \pm 0.04$, $G_{left}=0.13 \pm 0.04$; 
Complex: $G_{up}=0.928 \pm 0.005$, $G_{down}=0.928 \pm 0.005$, $G_{right}=0.928 \pm 0.006$, $G_{left}=0.928 \pm 0.006$; 
Chaotic: $G_{up}=0.9776 \pm 0.0025$, $G_{down}=0.9776 \pm 0.0025$, $G_{right}=0.9775 \pm 0.0025$, $G_{left}=0.9775 \pm 0.0025$.
}
\end{figure}

\begin{figure}
\hspace*{-0.4cm}
\includegraphics[scale=0.42]{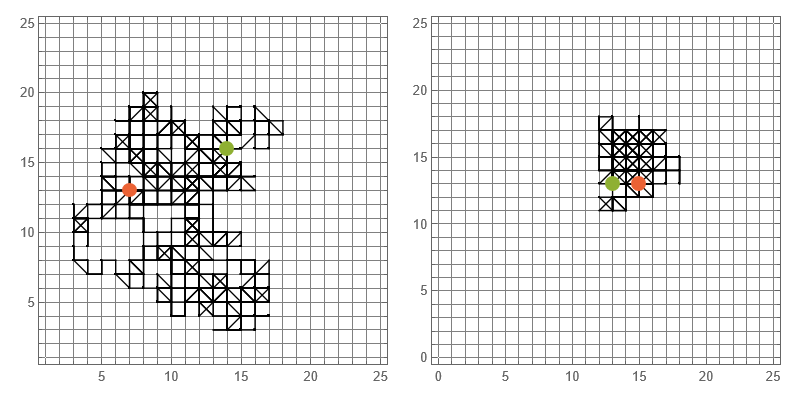}
\caption{\label{wide:rw} Trajectory of agent's average position in the complex (left) and chaotic (right) regimes, from initial (green) to final (orange) position. In the complex regime, the system behaves like a coordinated macro agent performing a two dimensional random walk, exploring the available space. In contrast, the chaotic regime lacks collective coordination and remains localized near its starting point, without a clear spatial trend.}
\end{figure}

\textit{Results.--}The results corresponding to the MIG for each type of regime behavior are presented in Fig.~\ref{wide:gconv}. In the convergent and periodic regimes, MIG remains consistently low throughout the entire evolution, with mean values over all directions $G=0.1192 \pm 0.0024$ and $G=0.135 \pm 0.020$ respectively. Their similarity may arise because, under the MIG, oscillatory motion does not necessarily imply spatial redistribution. The higher standard deviation observed in the periodic regime could result from the formation of multiple clusters oscillating in different directions, unlike the convergent regime where all agents collapse into a single point. In contrast, complex and chaotic regimes produce significantly higher MIG, with mean values over all directions $G=0.9279 \pm 0.0027$ and $G=0.9776 \pm 0.0012$ respectively. The complex regime exhibits a lower MIG due to agents with coordinated movement, but their similarity may arise because agents in both regimes remain confined within a reduced spatial region and the distinction in their spatial distributions is less perceptible to the MIG. To show that our metric captures meaningful differences, we further analyzed the regimes with similar MIG, showing clear distinctions in their dynamics. For complex and chaotic regimes, we examined the mean agent positions over time. As shown in Fig.~\ref{wide:rw}, the complex regime resembles a coordinated two-dimensional random walk, with agents collectively exploring the space, whereas the chaotic regime lacks discernible structure and remains localized near the initial configuration. Similarly, although both the periodic and convergent regimes exhibit low MIG, only the periodic regime maintains nonzero positional variance near the end of the simulation, indicating sustained movement. In contrast, the convergent regime becomes fully static, with agents reaching fixed positions and zero variance across all axes.\\

\textit{Concluding remarks.--} We implemented a multi-agent, rule-based biased random walk model capable of reproducing the four behavioral classes described by S. Wolfram: \textit{convergent, periodic, complex}, and \textit{chaotic}. The complex regime displays an emergent behavior in which agents self-organize into a macro-agent that collectively moves resembling a two-dimensional random walk. We employ the Mean Information Gain (MIG), a conditional entropy–based metric that quantifies the lack of information about other elements of a structure, effectively measuring the reduction of the average uncertainty between consecutive spatial configurations. This measure captures the degree of spatial disorder across regimes and provides a natural way to order them from low to high disorder as convergent, periodic, complex, and chaotic. The results confirm that the measure effectively distinguishes ordered behavior from disordered behavior. This provides a reliable quantitative basis for identifying and comparing emergent behaviors that can be readily extended to larger systems. Low MIG in the convergent and periodic regimes reflect their ordered nature, characterized by minimal spatial redistribution of agents. The slight difference observed in terms of variance in the periodic regime probably arises from the presence of multiple clusters oscillating in different directions rather than complete convergence to a single point. In contrast, complex and chaotic regimes produce substantially higher MIG, revealing the metrics' ability to discern different forms of disorder. The complex regime exhibits coordinated exploration of space, whereas the chaotic regime remains uncoordinated and spatially confined. The numerical similarity observed across both pairs of regimes can be attributed to the spatial confinement imposed by the model and the limited number of occupancy states (occupied or unoccupied), which constrain the resolution of spatial differentiation captured by the MIG. Together, these results reinforce the generality of MIG as a measure of structural complexity, supporting its relevance not only in physical systems but also in broader contexts such as ecological organization \cite{PARROTT20101069}.

\textit{Future work.--} Our toy model exhibits rich and non-trivial behavior, particularly in the complex regime, suggesting that further investigation could provide deeper insights into its dynamics. The emergence of a macro-agent exhibiting a coordinated random walk deserves closer examination. Future work could aim to characterize the type of random walk performed, explore its properties in higher dimensions or alternative spatial topologies, and draw connections to known collective behaviors, such as flocking dynamics. It would also be valuable to identify real-world systems or environments in two-dimensional space that exhibit similar emergent motion. In addition, the model can be extended by treating the \textit{vision} parameter as continuous rather than binary, allowing smooth transitions between behavioral regimes. Another possible extension involves setting a maximum number of agents per patch to study stacking effects. To better characterize these stacked configurations, the MIG could be computed over more complex states, such as the number of agents per cell, providing a finer-grained measurement of spatial complexity and local order. Another promising direction is to compare MIG with measures derived from the Kolmogorov algorithmic complexity \cite{javaheri2016information}, to test whether it can similarly distinguish the four Wolfram behavioral classes.

\bibliography{apssamp}

\end{document}